\documentclass[conference]{IEEEtran}
\usepackage{cite}
\usepackage{amsmath,amssymb,amsfonts}
\usepackage{algorithmic}
\usepackage{graphicx}
\usepackage{textcomp}
\usepackage{xcolor}

\usepackage{epstopdf}
\usepackage{enumerate}
\usepackage{subcaption}

\newcommand{\complex}{\mathbb{C}}
\renewcommand{\natural}{\mathbb{N}}

\newcommand{\bm}{\mathbf}

\newcommand{\delaymatrix}{\bm{\Pi}}
\newcommand{\dopplermatrix}{\bm{\Delta}}
\newcommand{\delayindex}{l}
\newcommand{\dopplerindex}{k}

\newcommand{\bL}{\mathbf{L}}

\newcommand{\subcarrier}{\Delta f}
\newcommand{\kron}{\otimes}

\begin{document}

\title{OTFS: Interleaved OFDM with Block CP\\
}

\author{\IEEEauthorblockN{Vivek Rangamgari*, Shashank Tiwari*, Suvra Sekhar Das*, and Subhas Chandra Mondal${}^\dagger$ \\}
	*G.S Sanyal School of Telecommunication, Indian Institute of Technology, Kharagapur, India\\
	${}^\dagger$ Wipro Limited, Bangalore, India
	
}

\maketitle

\begin{abstract}
Orthogonal time frequency space (OTFS) modulation is a recently proposed waveform for reliable communication in high-speed vehicular communication scenarios. It has better resilience to inter-carrier interference (ICI) than orthogonal frequency division multiplexing (OFDM). In this work, we describe OTFS as block-OFDM with a cyclic prefix and time interleaving. This interpretation helps one visualize OTFS in the light of OFDM as well as it also helps in analyzing the gain obtained by OTFS over OFDM. Further, we compare the performance of OTFS with its contender 5G new radio (NR)'s OFDM configuration of variable subcarrier bandwidth (VSB-OFDM) while considering practical forward error correction codes and 3GPP high-speed channel model. This provides realistic performance comparison, which is highly desired for technology realization. Considering practical channel estimation, we find that OTFS outperforms VSB-OFDM with 5G NR parameter by about 5dB. We also present results on peak to average power ratio (PAPR) due to specific pilot structure used in OTFS for channel estimation.
\end{abstract}


\section{Introduction}
There is an increasing demand for providing good quality of service in high speed vehicular scenarios \cite{liang_vehicular_2017} such as in vehicle to vehicle communications (V2V), unmanned aerial vehicle communications, etc in 5G. Orthogonal frequency division multiplexing (OFDM), which is a popular transmission technology, is limited in providing reliable connection in high speed vehicular scenarios. This is due to its high sensitivity to inter carrier interference (ICI), caused by Doppler spread and phase noise. Orthogonal time frequency space (OTFS) \cite{hadani_orthogonal_2017} is shown to be superior to OFDM in such high mobility environments, which is an integral part of 5G's operating scenarios.

In OTFS, the user data (constellation symbols) is placed in the delay-Doppler (De-Do) domain as opposed to time-frequency grid in OFDM. The data is then spread across the time-frequency grid using a unitary transform. This is followed by an OFDM \cite{hadani_orthogonal_2017} or block OFDM \cite{raviteja_practical_2019} modulator. Cyclic prefix (CP) \cite{hadani_orthogonal_2017} or block CP \cite{raviteja_practical_2019} is added to absorb the channel delay spread. When OFDM modulation is used with CP, it is called as CP-OTFS whereas, OTFS with block OFDM modulation and block CP is termed as reduced CP OTFS (RCP-OTFS) \cite{raviteja_otfs_2019}. In this paper, we consider the RCP-OTFS, which is spectrally more efficient than CP-OTFS.

The time-frequency spread OTFS signal, which suffers from inter symbol interference (ISI) and ICI in linear time varying (LTV) channel \cite{hadani_otfs:_2018}, is processed using advanced interference cancellation receiver. Such receivers can be of two types, namely (i) Linear receivers such as in \cite{tiwari_low_2019} and (ii) 
Non-linear receivers such as in \cite{raviteja_interference_2018}. Non-linear receivers have lower error probability but have higher computational complexity than linear receivers. We limit our work to linear receiver considering practical realizability.\\
\indent In order to deal with high ICI due to Doppler spread and phase noise, 5G NR has adopted a variant of contemporary edition of OFDM, which is variable subcarrier bandwidth OFDM (VSB-OFDM) \cite{das2005} \cite{das_variable_2007-1} \cite{das_multi_2005}. The VSB reconfigurability is expressed in terms of `numerology'. It is established that VSB-OFDM has higher resilience to ICI \cite{das_variable_2007}. Comprehensive comparison of VSB-OFDM with OTFS is important to select efficient waveform for the environments where Doppler and phase noise effects are strong.\\
\indent The gain of OTFS over VSB-OFDM is mainly attributed to diversity achieved by time-frequency spreading of signal\cite{hadani_otfs:_2018}. On the other hand, practical systems use forward error correction (FEC) codes which improve error performance by virtue of diversity introduced through redundancy and advanced singal processing at receiver. Thus, the overall diversity gain that coded OTFS has over coded VSB-OFDM needs to be studied, which is one of the objectives of this work.
 In \cite{hadani_otfs:_2018}, VSB-OFDM is compared with CP-OTFS. Uncoded performance of fixed subcarrier bandwidth OFDM is compared with RCP-OTFS\cite{raviteja_interference_2018} and CP-OTFS\cite{wiffen_comparison_2018}, whereas performance of coded RCP-OTFS with VSB-OFDM (5G NR) is yet to receive attention, which is provided in this work.\\
\indent The contributions in this paper can be summarized as,
\begin{itemize}
\item We show that OTFS can be interpreted as a block OFDM with a single CP and time interleaving of the samples of all OFDM symbols. This interpretation simplifies the way we understand OTFS signal generation thereby paving the path for a reconfigurable transceiver architecture.
\item Through this novel interpretation of OTFS, we establish that diversity gain in OTFS is due to time-interleaving. We also seggregate the gains due to the use of advanced interference cancellation receiver and time-interleaving.
\item In order to present realistic results for performance comparison, we evaluate the performance of both transmission technologies with
\begin{enumerate}[(i)]
\item channel profile specified in \cite{3gpp38901} by 3GPP.
\item practical channel estimation algorithms, and
\item LDPC codes
\end{enumerate}
\item The pilot structure used for channel estimation in OTFS is different from VSB-OFDM. We hypothesize that such pilot structure should affect the peak to average power ratio (PAPR) of OTFS, hence present the related results.
\end{itemize}

We follow the  notations described below throughout the paper. We use $\bm{Z}$, $\bm{z}$, $z$ as Matrix, Vector and scalar respectively. $()^{T}$ and $()^{\dagger}$ denote transpose and hermitian operations. $W_L$ and $I_N$ represents $L$ order normalized Inverse Discrete Fourier Transform(IDFT) matrix and $N$ order Identity matrix. Kronecker product operator is given by $\kron$. The operator $\mathrm{diag}\{\bm x\}$ creates a diagonal matrix with the  elements of vector $\bm x$.
Circulant matrix is represented by $circ\{\bm{ x}\}$  whose first column is $\bm{x}$. Notations $\lfloor{-}\rfloor$and $\lceil{-}\rceil$ are floor and ceil operators respectively. Column-wise vectorization of matrix $(\bm{ X})$ is represented by  $vec\{\bm{X}\}$  and $j=\sqrt{-1}$. $\natural[a~ b]$ denotes set of natural numbers between $a$ and $b$.

\section{RCP-OTFS}

\subsection{OTFS Transmission}

We consider RCP-OTFS system operating with time-frequency resource of total $T_f$ seconds duration and $B$ Hz. The Bandwidth is divided into $M$ number of sub-carriers having $\subcarrier$ sub-carrier bandwidth and we transmit $N$ number of symbols having $T$ symbol duration, thus $B=M\subcarrier$ and $T_f=NT$. Furthurmore, OTFS is critically sampled, i.e $T\subcarrier = 1$

The source bitstream is encoded using LDPC codes and then passed through symbol mapper. The QAM modulated data and pilot symbols are arranged over De-Do lattice $\Lambda=\{(\frac{k}{NT},~\frac{l}{M \subcarrier})\}$, $k\in \natural[0~N-1]$, $l\in \natural[0~M-1]$ as shown in Fig.\ref{otfs_pilot}. De-Do signal can be given as,
\begin{equation}
x(k,l) = \begin{cases}
x_p, & k = K_p~ \& ~l=L_p \\
0 & K_p-2k_{\nu} \leq k \leq K_p + 2k_{\nu} ~ \& ~ \\
& L_p-2l_{\tau} \leq l \leq L_p + 2l_{\tau}\\
d(k,l), & otherwise
\end{cases}
\end{equation}

where $d(k,l) \in \complex$ is the QAM data symbol. We assume that $E[d(k,l) \bar{d}(k',l')] = \sigma_d^2 \delta(k-k',l-l') $, where $\delta$ is Dirac delta function. $x_p = \sqrt{P_{plt}}$ is the pilot symbol, which is a 2-D discrete impulse at location $(K_p,L_p)$, where $ K_p \in \natural[2k_{\nu}+1 ~ N-2k_{\nu}-2],~ L_p \in \natural[l_{\tau}+1 ~ M-l_{\tau}-2]$ and $P_{plt}$ is power of pilot symbol. $k_{\nu} = \lceil\nu_{max}NT\rceil$ and $l_{\tau} = \lceil\tau_{max}M\subcarrier\rceil$ are maximum doppler length and delay length of the channel\cite{raviteja_embedded_2019}.\\
\indent $x(k,l)$ is mapped to time-frequency data $Z(n,m)$  on lattice $\Lambda^{\perp}=\{(nT,~m\subcarrier)\}$, $n\in \natural[0~N-1]$ and $m\in \natural [0~M-1]$ by using inverse symplectic finite Fourier transform (ISFFT) as shown in Fig.1 of \cite{raviteja_interference_2018}. Thus $Z(n,m)$ can be given as,

\begin{equation} \label{isfft}
Z(n,m)=\frac{1}{\sqrt{NM}} \sum_{k=0}^{N-1}{\sum_{l=0}^{M-1}{x(k,l) e^{j2\pi [\frac{nk}{N}-\frac{ml}{M}]}}}.
\end{equation}

The time domain signal is obtained from $Z(n,m)$ using Heisenberg transform as,
\begin{equation}
s(t)= \sum_{n=0}^{N-1}{\sum_{m=0}^{M-1}}{Z(n,m)g(t-nT) e^{j2\pi m \subcarrier (t-nT)}}
\end{equation}
where, $g(t)$ is transmitter pulse of duration $T$. In this paper, we consider $g(t)$ to be a rectangular pulse as in \cite{raviteja_practical_2019},i.e,
\begin{equation}
g(t) = \begin{cases}
1 & 0 \leq t \leq T\\
0 & \mathrm{otherwise}
\end{cases}
\end{equation}
To denote the above system in its equivalent matrix-vector form, we obtain the discrete version of OTFS system by sampling $s(t)$ at the sampling interval of $\frac{T}{M}$ and  $\bm{s}=[s(0)~s(1) \cdots s(MN-1)]$ is formed from the samples of $s(t)$.\\
\indent If the De-Do symbols $x(k,l)$ are arranged in $M\times N$ matrix as,
\begin{equation}
\mathbf{X}= \small{\begin{bmatrix}
	x(0,0) & x(0,1) & \cdots & x(N-1,0) \\
	x(0,1) & x(1,1) & \cdots & x(N-1,1) \\
	\vdots & \vdots & \ddots & \vdots  \\
	x(0,M-1) & x(1,M-1) & \cdots & x(N-1,M-1) \\
	\end{bmatrix}}
\end{equation}
Then, the time domain signal can also be written as matrix-vector multiplication,
\begin{equation} \label{Eqn:OTFSTxSig}
\bm{s}= \bm{A} \bm{x}
\end{equation}
where, $\bm{x}=vec(\bm{X})$ and $\bm{A}= \bm{W_N} \kron \bm{I_M} $ denotes the OTFS transform ISFFT matrix. A cyclic prefix (CP) of length $L \geq l_{\tau}$ is appended at the start of the $\bm{s}$.

\subsection{OTFS as interleaved Block OFDM}\label{otfs_bofdm}
\begin{figure}[h] \label{interl_fig}
	\centerline{\includegraphics[width=\linewidth]{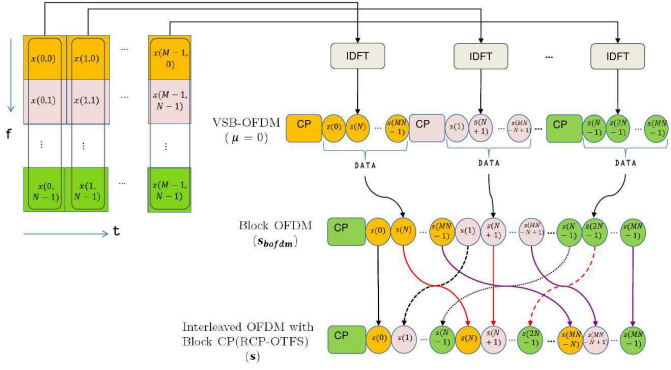}} 
	\caption{Time domain signal construction in CP-OFDM, Block OFDM and RCP-OTFS}
	\label{fig:otfs_time}
\end{figure}

The time-frequency signal obtained in (\ref{isfft}) can be expresesed in matrix form $\bm{Z}$ as,
\begin{equation}
\bm{Z} = \bm{W_M}^{\dagger} \bm{X} \bm{W_N}
\end{equation}
The OTFS time domain signal $\bm{s}$ is obtained by applying IDFT w.r.t frequency domain and written as,
\begin{equation}
\bm{s} = vec\{\bm{W_M}\bm{Z}\}
\end{equation}
which can be simplified as,
\begin{equation} \label{interleaved_ofdm}
\bm{s} = vec\{\bm{X} \bm{W_N}\}
\end{equation}
Moreover, if Block OFDM is used to transmit the same data ($\bm{X}$) assuming that it is in the time-frequency domain with $N$ subcarriers and $M$ OFDM symbols, then the signal $\bm{s_{bofdm}}$ in time domain can be expressed as,
\begin{equation} \label{ni_ofdm}
\bm{s_{bofdm}} = vec\{\bm{W_{N}} \bm{X}^{T}\}
\end{equation}
From (\ref{interleaved_ofdm}) and (\ref{ni_ofdm}), 
\begin{equation} \label{int_relation}
\bm{s}(nM+m) = \bm{s_{bofdm}}(mN+n)
\end{equation}
where  $m \in \natural[0~M-1]$ and $n \in [0~N-1]$.\\
Therefore, it can be said that transmitted OTFS signal $\bm{s}$ can be obtained by shuffling transmitted block OFDM signal $\bm{s_{bofdm}}$ using the simple relationship as in (\ref{int_relation}). Thus, it is established that OTFS can be seen as block OFDM with time interleaving.
Moreover, it must be carefully noted that parmeters for OTFS and its equivalent Block OFDM system are different. For the same time-frequency resource of $T_f$ sec and $B$ Hz, OTFS system has M subcarriers and N OTFS symbols, whereas equivalent Block OFDM system has $M_{bofdm} = N$ subcarriers and $N_{bofdm}=M$ OFDM symbols. Therefore, the subcarrier bandwidth for the equivalent Block OFDM sytem changes to $\subcarrier_{bofdm} = \frac{M}{N}\subcarrier$ and OFDM symbol duration changes to $T_{bofdm} = \frac{N}{M} T$.
In general $M > N$, thus $\subcarrier_{bofdm} > \subcarrier$ which implies that OTFS will have increased capability to combat Doppler as opposed to an OFDM system having $\subcarrier$ sub-carrier bandwidth. Fig.~\ref{fig:otfs_time} shows the time domain signal generation for different waveforms and depicts the RCP-OTFS as block OFDM with time interleaving.

\subsection{Channel}\label{OTFSchannel}
We consider a time varying channel of $P$ paths\cite{tiwari_low_2019} with $h_{p}$ being the complex coefficient at $\tau_{p}$ delay and $\nu_{p}$ Doppler frequency of the $p$th path where $p \in \natural[1~P]$. Thus, De-Do channel spreading function can be given as,
\begin{equation}
h(\tau,\nu)=\sum_{p=1}^{P}{h_{p} \delta(\tau-\tau_{p}) \delta(\nu-\nu_{p})}
\end{equation}
The delay and Doppler values for $p$th path is given as $\tau_{p}=\frac{\delayindex_p}{M\subcarrier}$ and $\nu_{p}=\frac{\dopplerindex_p}{NT}$ where $\delayindex_p\in \natural[0~M-1]$ and $\dopplerindex_p\in\natural[0~N-1]$ are delay and Doppler bin number on De-Do lattice $\Lambda$ for $p^{\rm th}$ path.  In this work, we assume that $N$ and $M$ are sufficiently large so that there is no effect of fractional delay and Doppler on the performance\cite{raviteja_interference_2018}. \\
We also define time varying frequency response of channel as,
\begin{equation}
h_{tf}(f,t) = \int_{0}^{\tau_{max}}\int_{-\nu_{max}}^{\nu_{max}}h(\tau,\nu)e^{j2\pi(\nu t- f\tau)} d\nu d\tau
\end{equation}
which simplifies as,
\begin{equation}
h_{tf}(f,t) = \sum_{p=1}^{P} h_{p}e^{j2\pi(\nu_p t- f\tau_p)}
\end{equation}
Its discrete version $\acute{h}(m,n)$, the channel coefficient at $m$th subcarrier of $n$th time symbol, used later in \ref{ofdm_rec}, is defined as,
\begin{equation} \label{mn_coeff_ch}
\acute{h}(m,n) = h_{tf}(f,t) \bigg|_{f = m\subcarrier,t = nT}
\end{equation}

\subsection{Receiver} \label{OTFSreceiver}
After removal of CP at the receiver, the received signal can be written as \cite{tiwari_low_2019},
\begin{equation}
\bm{r} =\bm{H} \bm{s} +\bm{n}
\end{equation}
where, $\bm{n}$ is white Gaussian noise vector of length $MN$ with elemental variance $\sigma_{\bm{n}}^2$ and $\bm{H}$ is a  $MN\times MN$ channel matrix and can be given as,
\begin{equation}
\bm{H} =\sum_{p=1}^{P}{h_p \delaymatrix^{\delayindex_p} \dopplermatrix^{\dopplerindex_p}}
\end{equation}
with $\delaymatrix=circ\{[0~1~0 \cdots 0]^{T}_{MN\times 1}\}$ is a circulant delay matrix and $\dopplermatrix=diag\{[1~ e^{j2\pi\frac{1}{MN}}~ \cdots e^{j2\pi\frac{MN-1}{MN}}]^{T}\}$ is a diagonal Doppler matrix.\\

\begin{figure}[t] 
	
\begin{subfigure}[b]{0.85\linewidth}
	\centering
			\includegraphics[scale=0.17]{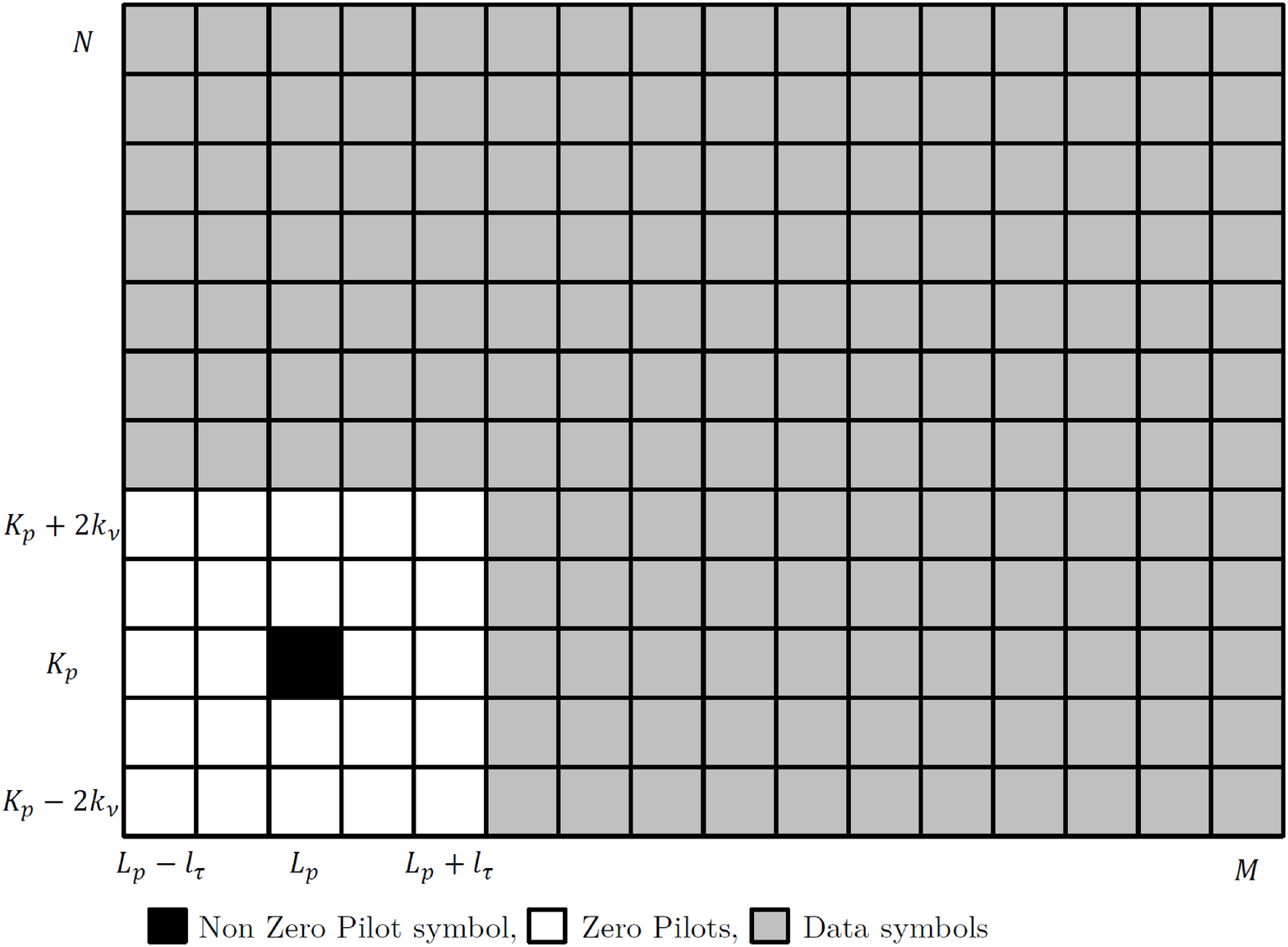}
		\caption{RCP-OTFS}
		\label{otfs_pilot}

\end{subfigure}
\hfill
\begin{subfigure} [b]{0.9\linewidth}
	\centering
	\includegraphics[scale=0.14]{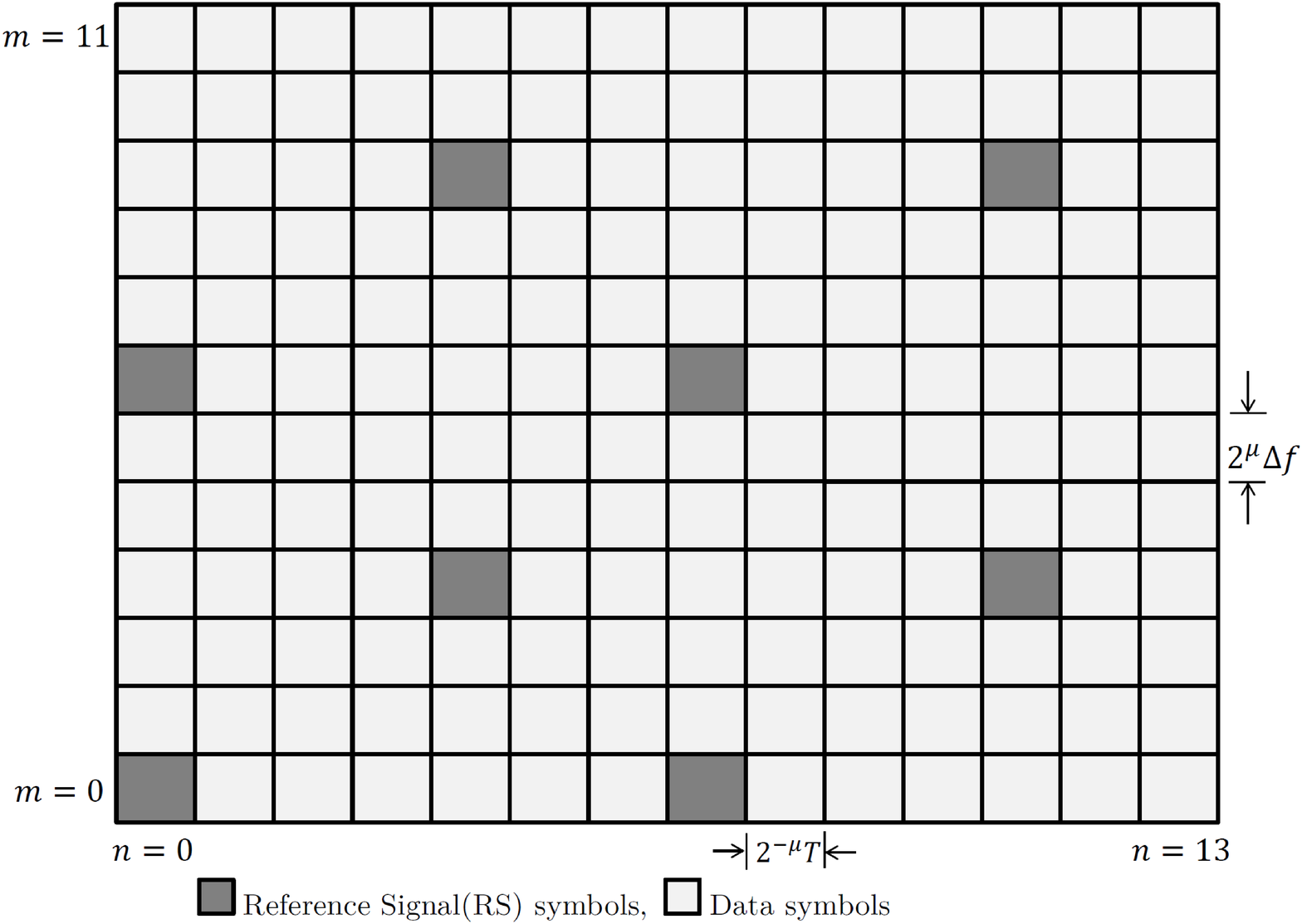}
	\caption{VSB-OFDM}
	\label{ofdm_prb}
	\end{subfigure}
\caption{Pilot placement in RCP-OTFS and in a PRB of VSB-OFDM }
\label{pilots_fig}
\end{figure}

\paragraph{Channel Estimation}
Channel matrix $\bm{H}$ is estimated using the pilot symbols. 
The received signal is transmformed to De-Do domain as,
\begin{equation}
\bm{y} = \bm{A}^{\dagger} \bm{r}
\end{equation}
The received vector $\bm{y}$ is then reshaped into $N \times M$ grid as
\begin{equation}
\bm{Y}(k,l) = \bm{y}(k+lN)
\end{equation}
where $~ k \in \natural[0~ N-1],~ l \in \natural[0~ M-1]$.
The channel estimation from here proceeds according to \cite{raviteja_embedded_2019}, in which the non zero pilot at location $(K_p,L_p)$ as shown in Fig.\ref{otfs_pilot} is spread to locations $(k,l), k \in \natural[K_p - k_{\nu}~~K_p + k_{\nu}]$ and $l \in \natural[L_p~~L_p + l_{\tau}]$ because of the channel.

The channel parameters $(h_p,k_p,l_p)$ are extracted from this region using the threshold based scheme with threshold ($\Upsilon$) as $3\sigma_{\bm{n}}$ as described in \cite{raviteja_embedded_2019}. Then,
\begin{equation}
\begin{split}
(\hat{h}_p,\hat{k}_p,\hat{l}_p) = (Y(k,l),k,l) ~
\ni \| Y(k,l) \| > \Upsilon \\ ~ \forall ~ K_p-k_{\nu} \leq k \leq K_p + k_{\nu} ~ \& ~  L_p \leq l \leq L_p + l_{\tau}
\end{split}
\end{equation}
where $p \in \natural[1 ~\hat{P}]$ and $\hat{P}$ is the number of taps detected.\\
The estimate of matrix $\bm{H}$ is then given as,
$$\hat{\bm{H}} = \sum_{p=1}^{\hat{P}}{\hat{h}_p \delaymatrix^{\hat{\delayindex}_p} \dopplermatrix^{\hat{\dopplerindex}_p}}$$

\paragraph{Equalization}
The De-Do data is estimated by equalizing the received time domain signal using MMSE \cite{tiwari_low_2019} as,
\begin{equation}
\hat{\bm{x}}= \bm{H_{mmse}} \bm{r}
\end{equation}
where $\bm{H_{mmse}} = (\bm{\hat{H}}\bm{A})^{\dagger} [(\bm{\hat{H}} \bm{A})(\bm{\hat{H}} \bm{A})^{\dagger}+ \frac{\sigma_n^2}{\sigma_d^2} \bm{I}]^{-1}$.

To decode the equalized data using LDPC decoder, the log-likelihood ratios (LLRs) are passed to the decoder which are calculated from the equalized symbols as,
\begin{equation} \label{LLR_eqn}
LLR(b_{\eta}^j | \hat{x}(\eta)) \approx ({\min_{s\epsilon S_{j}^{0}}}\frac{||\hat{x}(\eta)-s||^{2}}{\bm{\sigma^2}(\eta,\eta)}) - ({\min_{s\epsilon S_{j}^{1}}}\frac{||\hat{x}(\eta)-s||^{2}}{\bm{\sigma^2}(\eta,\eta)})
\end{equation}
where $\bm{\hat{x}}(\eta)$ is the $\eta^{th}$ element of $\bm{\hat{x}}$ mapped from the bits $b_{\eta}^{0}~b_{\eta}^{1}\cdots b_{\eta}^{K-1}, K$ is the number of bits per symbol and $\boldsymbol{\sigma^2}(\eta,\eta)$ is the element of $\boldsymbol{\sigma^2} = \sigma_n^2(\bm{H_{mmse}} \bm{H_{mmse}^{\dagger}})$. $S_j^0$ and $S_j^1$ denotes the set of constellation symbols where the bit $b^j_{\eta} = 0$ and $b^j_{\eta} = 1$ respectively for $j=0,1,\cdots,K-1$.

These LLRs are then fed into the LDPC decoder to decode data.
Let $\bm{L}$ denotes a matrix where $\bL(\eta,j) = LLR(b_{\eta}^j | \bm{\hat{x}(\eta))}$ for $\eta = 1,2,\cdots,MN$ and $j=0,1,\cdots,K-1$. 
$\bL$ is reshaped to $L_{cl}\times N_{cw}$ matrix where $L_{cl}$ and $N_{cw}$ denote the LDPC codeword length and number of codewords respectively.
Each column of $\bL$ subsequently regenerates message word $m_{\iota}$ for $\iota=1,2\cdots,N_{cw}$ using the Min-Sum algorithm\cite{Zhao_minsum} employed by the LDPC decoder and is collected as the recovered data.

\section{VSB OFDM Overview}
We consider a VSB-OFDM system with total frame duration $T_f$ sec. and bandwidth $B$ Hz. 
We have a base sub-carrier bandwidth of $\subcarrier$ Hz and variability in the sub-carrier bandwidth is introduced by the parameter $\mu$. For a given $\mu$, the subcarrier bandwidth is $2^{\mu} \subcarrier$ Hz. We have total $2^{-\mu}M$ number of sub-carriers and $2^{\mu}N$ number of OFDM symbols having $2^{-\mu}T$ symbol duration , thus $B=M\subcarrier$ and $T_f=NT$.\\

\subsection{Transmitter}
We use the Physical Resource Block (PRB) frame structure of 5G NR for VSB-OFDM system. Each PRB consists of $12$ subcarriers $\times$ $14$ time slots with $8$ reference signals (RS) as shown in Fig.\ref{ofdm_prb}. Then the PRBs are arranged to fill the time-frequency grid of $B$ Hz and $T_f$ sec. The number of PRBs in a given frame is,
$$ N_{PRB} = \left\lfloor\frac{B}{12.2^{\mu}.\subcarrier}\right\rfloor \left\lfloor \frac{T_f}{14.2^{-\mu}.T}\right\rfloor $$
The source bits are encoded using LDPC code and then passed through the symbol mapper. The modulated QAM symbols are arranged in the $N_{PRB}$ number of PRBs to form the time-frequency signal $\acute{x}(m,n)$, $m \in \natural[0 ~~ 2^{-\mu}M-1]$ and $n \in \natural[0~~2^{\mu}N-1]$ in a frame. Then, the $n$th time domain OFDM symbol can be given as,
\begin{equation}
\acute{s}_n(t) = \sum_{m=0}^{2^{-\mu}M-1} \acute{x}(m,n)e^{j2\pi m 2^{\mu}\subcarrier (t-n2^{-\mu}T)}
\end{equation}
where $2^{-\mu} T$ is the one OFDM symbol duration without CP. 
Further, if $\bm{\acute{x}_n}$ represents the discrete version of $\acute{s}_n(t)$ sampled at $\frac{1}{B}$, then the $2^{\mu}N$ concatenated OFDM symbols can be given as,
\begin{equation}
[\bm{\acute{x}_0} ~\bm{\acute{x}_1}~ \cdots ~\bm{\acute{x}_{2^{\mu}N}}] = \bm{W_{2^{-\mu}M}} \bm{\acute{X}}
\end{equation}
where $\bm{\acute{X}}$ is the time-frequency frame given as,
$$
\mathbf{\acute{X}}= \small{\begin{bmatrix}
	\acute{x}(0,0) & \cdots & \acute{x}(0,2^{\mu}N-1) \\
	\acute{x}(1,0)  & \cdots & \acute{x}(1,2^{\mu}N-1) \\
	\vdots & \ddots & \vdots  \\
	\acute{x}(2^{-\mu}M-1,0) & \cdots & \acute{x}(2^{-\mu}M-1,2^{\mu}N-1) \\
	\end{bmatrix}}
$$
Then, Cyclic Prefix (CP) is appended to each OFDM symbol of duration $2^{-\mu}T_{cp}$ sec such that $T_{o} = T + T_{cp}$\\

\subsection{Receiver} \label{ofdm_rec}
We assume that the CP duration is greater than maximum excess delay of channel and OFDM system's symbol duration and subcarrier bandwidth are less than coherence time and coherence bandwidth of the channel respectively. Then, after removal of CP and applying DFT on the residual signal, the received time-frequency symbol $\acute{y}(m,n)$ can be expressed as,
\begin{equation}
\acute{y}(m,n) = \acute{h}(m,n)\acute{x}(m,n) + \acute{v}(m,n)
\end{equation}
where, $m \in \natural[0 ~~ 2^{-\mu}M-1]$  is the subcarrier index and $n \in \natural[0~~2^{\mu}N-1]$ is the OFDM symbol index. $\acute{h}(m,n) \in \complex$ defined in (\ref{mn_coeff_ch}) is the channel coefficient and $\acute{v}(m,n) \in \complex$ is white guassian noise with variance $\sigma_{V}^2$ at $m$th subcarrier  of $n$th OFDM symbol.

For channel estimation, the RS in the PRB are used to obtain the estimates at the pilot location using MMSE estimation as,
\begin{equation}
\hat{\acute{h}}(m_{RS},n_{RS}) = \frac{\acute{x}(m_{RS},n_{RS})^{\dagger} ~ \acute{y}(m_{RS},n_{RS})}{\|\acute{x}(m_{RS},n_{RS})\|^2 + \sigma_{V}^2}
\end{equation}
where $(m_{RS},n_{RS}) = \{(0,0),(6,0),(3,4),(9,4),(0,7),(6,7),\\(3,11),(9,11)\}$ are the locations of RS in a PRB.

The obtained channel estimates at the RS locations in PRBs are interpolated to get the channel estimates at the data locations using DFT interpolation along frequency axis and linear interpolation along time axis as described in \cite{fernandez_DFT_ch}. \\
The estimate of data symbols are obtained at the receiver as,
\begin{equation}
\hat{\acute{x}}(m,n) = \frac{\acute{y}(m,n)}{\hat{\acute{h}}(m,n)} 
\end{equation}
The estimated symbols are used to generate the channel LLR values of bits corresponding to the symbol by substituting $\boldsymbol{\sigma^2} = diag\{vec\{\boldsymbol{\sigma_{V_{eff}}^2}\}\}$ in (\ref{LLR_eqn}) where $\sigma_{V_{eff}}^2(m,n) = \frac{\sigma_V^2}{\|\hat{\acute{h}}(m,n)\|^2}$. Then, the LLRs are used to recover the data as described in \ref{OTFSreceiver}.

\section{Pilot Power in OTFS and VSB-OFDM}
 In this section we describe the pilot structure to be used in OTFS as in \cite{raviteja_embedded_2019}. We also describe the pilot structure used for evaluating VSB-OFDM. In the performance evaluation, we intend to keep same total transmit power for OTFS and OFDM. As shown in Fig.\ref{ofdm_prb}, there are 8 pilots per PRB in VSB-OFDM system considered. The total number of pilots in VSB-OFDM frame is $N_{p,ofdm} = 8N_{PRB}$, while total number of pilots in RCP-OTFS is $N_{p,otfs} = (4k_{\nu}+1)(2l_{\tau}+1)-1$\cite{raviteja_embedded_2019}. If we let the total pilot power to be equal, i.e $P_{plt} = n_{p,ofdm}P_{T} = n_{p,otfs}P_{T}$ where $P_{T}$ is the transmit power, $n_{p,otfs}$ and $n_{p,ofdm}$ are the ratio of number of pilot symbols to total number of symbols for RCP-OTFS and VSB-OFDM respectively. Since, RCP-OTFS uses only one non-zero pilot at location $(K_p,L_p)$  and ($N_{p,otfs}-1$) zero pilots as shown in Fig.\ref{otfs_pilot}, total pilot power is placed on the pilot symbol ($x_p$) which results in an uneven power distribution for pilots and data in RCP-OTFS, given by $\Delta P = 10log_{10}(\frac{P_{pilot}}{P_{data}})$ in De-Do domain. If $P_T$ is to be kept same for RCP-OTFS and VSB-OTFS systems with parameters given in Table \ref{sim_params}, then the value of $\Delta P$ is 34 dB. We evaluate the impact of $\Delta P$ on PAPR in Sec.\ref{results} below.
 

\section{Results} \label{results}
\begin{table}[htbp] \label{sim_params}
	\caption{Simulation Parameters}
	\begin{center}
		\begin{tabular}{|c|c|}
			\hline
			\textbf{Parameter} & \textbf{Value}\\
			\hline
			Carrier Frequency($f_c$) & 6 GHz \\
			\hline
			Bandwidth(B) & 7.68 MHz \\
			\hline
			Frame Time($T_f$) & 10 ms \\
			\hline
			Subcarrier Bandwidth($\subcarrier$) & 15 KHz\\
			\hline
			RCP-OTFS parameters & M=512, N=128, $K_p$=80, $L_p$=16\\
			\hline
			VSB-OFDM parameters & M=512, N=128, $\mu$=0,1,2,3 \\
			\hline
			Equivalent OFDM parameters & $M_{bofdm}$=128, $N_{bofdm}$=512,\\
		     &  $\subcarrier_{bofdm}$ = 60 KHz\\
			\hline
			Channel Model\cite{3gpp38901} & TDL-A, DS=37 ns, Rural Macro \\
			\hline
			CP duration($T_{cp}$) & 4.69 $\mu$s \\
			\hline
			UE speed & 500 kmph\\
			\hline
			FEC & QC-LDPC, coderate 2/3,\\
			&  Block Length 1944\\
			\hline
		\end{tabular}
		\label{sim_params}
	\end{center}
\end{table}

In this section, we present the LDPC coded performance of OTFS, block OFDM and VSB-OFDM system. The simulation parameters are mentioned in Table \ref{sim_params}. The key performance indicator (KPI) used to evaluate performance is block error rate (BLER), where block is coded using LDPC codes. 
For block OFDM system, we consider same frame size as RCP-OTFS. Time-frequency slots for block OFDM can be evaluated as already described in Sec.\ref{otfs_bofdm}, thus $M_{bofdm}=128$, $N_{bofdm}=512$, and $\subcarrier_{bofdm}$ = 60 KHz. It may be noted that we use the same LMMSE based interference cancellation equalizer for RCP-OTFS and block OFDM so that any difference in BLER performance can be attributed to interleaving only. The SNR losses due to CP, which are given as $10log_{10}(\frac{\lceil T_{cp}B\rceil}{M}) = 0.3$ dB for VSB-OFDM and $10log_{10}(\frac{\lceil T_{cp}B\rceil}{MN}) = 0.0023$ dB for OTFS. These SNR losses are also adjusted in the results. Moreover, same transmit power is ensured for all the transmission schemes.
 Doppler is generated following Jakes spectrum. The CP is chosen long enough to accommodate the maximum excess delay of the channel.

\begin{figure}[htbp]
	\centerline{\includegraphics[width=\linewidth ]{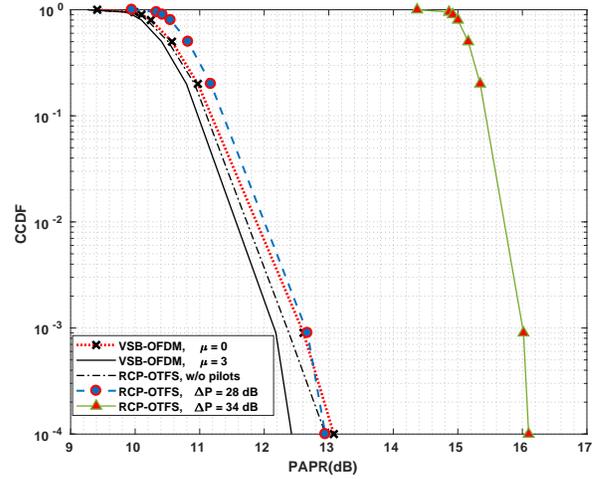}}
	\caption{CCDF of PAPR for OFDM and OTFS}
	\label{papr_result}
\end{figure}

\begin{figure}[htbp]
	\centerline{\includegraphics[width=\linewidth ]{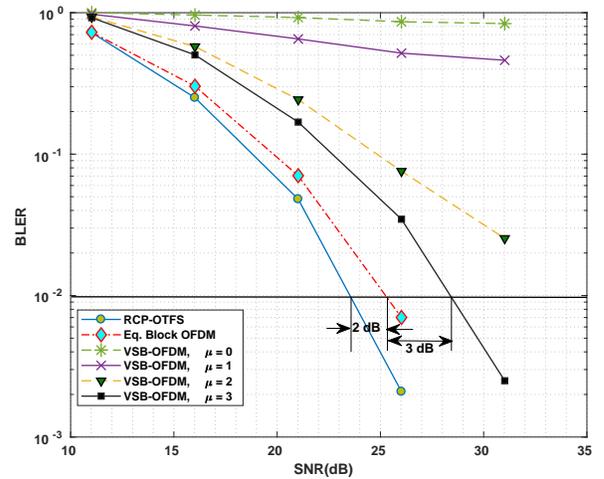}}
	\caption{BLER performance for 16QAM at 500kmph}
 \label{otfs_16qam}
\end{figure}

We hypothesize that a large $\Delta P$ may have an effect on PAPR of $\bm{s}$ \eqref{interleaved_ofdm}. We present the CCDF of RCP-OTFS for $\Delta P = 34$ dB and $\Delta P = 28$ dB in Fig.\ref{papr_result}. At $\Delta P = 34$ dB, which arises when we consider same $P_T$ for VSB-OFDM and RCP-OTFS. It can be observed that PAPR of RCP-OTFS is nearly 3.5 dB higher than that of VSB-OFDM ($\mu = 0$). We therefore reduce $\Delta P$ so that PAPR may be reduced. We have found that for given configuration, if $\Delta P \leq 28$ dB, there is no further reduction in PAPR of RCP-OTFS signal $\bm{s}$ than that of OFDM, and it is very close to that of VSB-OFDM with $\mu = 0$. Therefore, we have kept the $\Delta P =28$ dB in the performance evaluation presented here. Further, the reduction in $\Delta P$ implies that power of data RCP-OTFS is higher than that of VSB-OFDM by 0.154dB/symbol. Therefore, there is an additional gain in SNR for OTFS.

Figure \ref{otfs_16qam} presents the performance of different waveforms  for 16 QAM modulation at the vehicular speed of 500 kmph.  It can be observed that performance of VSB-OFDM becomes better with increasing value of $\mu$  as expected and supported by literature. This is because the effect of Doppler spread reduces with increased sub-carrier bandwidth as post processing SNR ($\Gamma$) for OFDM under Doppler, $ \propto (\frac{\subcarrier}{\nu_{max}})^2$ \cite{das_performance_2008}. It can also be observed that OTFS outperforms  VSB-OFDM even with $\mu=3$.\\
\indent At BLER of $10^{-1}$, OTFS has a SNR gain of 3.5 dB over VSB-OFDM ($\mu = 3$). This gain further increases to nearly 5 dB at BLER of $10^{-2}$. This result can be attributed to two reasons (i) OTFS has an interleaving gain  and, (ii) OTFS uses ICI cancellation LMMSE receiver whereas VSB-OFDM uses a simple single-tap receiver. 
Block OFDM has a SNR gain of around 2.5 dB and 3 dB at BLER $10^{-1}$ and $10^{-2}$ respectively. This gain is due to the use of interference cancelling receiver in block OFDM.
Further, at the BLER of $10^{-1}$ and $10^{-2}$, OTFS has gain of 1 dB and  2 dB respectively over block OFDM. As both block OFDM and RCP-OTFS use FEC and LMMSE equalizer while the difference in the  trasmitted signal is the time-interleaving
, henceforth it can be said that this gain is primarily due to the interleaving. 
We have observed similar results for other vehicular speeds as well as for other modulation schemes and chose to not include them here for brevity. Thus, it can be concluded that OTFS performs better than other contender waveforms in vehicular scenarios by virtue of its interleaving and LMMSE interference cancellation. 

\section{Conclusions}
 In this work, we have detailed the signal generation of OTFS and explained how it could be viewed in an alternative form as block OFDM with time-interleaving. From performance comparison results, we conclude that block OFDM outperforms VSB-OFDM by 3 dB due to the use of interference cancelling receiver. Inspite of using the same receiver as block OFDM, OTFS outperforms block OFDM. The only difference being the time-interleaving of the transmitted signal which provides an additional gain of 2 dB. Overall, OTFS has 5 dB gain over VSB-OFDM for 16 QAM in LTV channel with FEC.

We have also found that the pilot power of OTFS can be made lower than OFDM, since the channel estimation in OTFS is less sensitive to  estimation error. Furthermore, the percentage resource used for pilots and CP overhead in OTFS is less than OFDM, and hence the spectral efficiency (SE) of OTFS is higher than OFDM.
Finally, it can be said that due to OTFS's high SE, better resilience to ICI because of interference canceling receiver and interleaving, which is not usually available in an OFDM system with single tap equalizer, OTFS may be recommended as transmission technology for future generations of wireless communication systems.

\bibliographystyle{IEEEtran}
\bibliography{OTFS,OFDM,manual,HighDopplerCommunication}
 
\vspace{12pt}

\end{document}